%% file: IRF_three_block.tex
\input phyzzx\input mydef

\def\pr{\prime}
\par
\overfullrule=0pt

\date{July,  2018}
\date{July, 2018}
\titlepage
\title{Three Blocks Solvable Lattice Models and Birman--Murakami--Wenzl Algebra}
\author{Vladimir  Belavin$^{a,b,c}$ and Doron Gepner$^a$}
\vskip20pt
\line{\it\hfill  $^a$\  Department of Particle Physics and Astrophysics, Weizmann Institute, \hfill}
\line{\it\hfill Rehovot 76100,  Israel\hfill} 
\line{\it \hfill $^b$ \  I.E. Tamm  Department of Theoretical Physics, P.N. Lebedev Physical
\hfill }\line{\hfill\it  Institute, Leninsky av. 53, 11991 Moscow, Russia\hfill}
\line{\it \hfill $^c$ \   Department of Quantum Physics, Institute for Information Transmission\hfill}
\line{\hfill\it  Problems, Bolshoy Karetny per. 19,  127994 Moscow, Russia\hfill}

\abstract
Birman--Murakami--Wenzl (BMW) algebra was introduced in connection with knot theory. We treat here
interaction round the face solvable (IRF) lattice models.
We assume that the face transfer matrix obeys a cubic
polynomial equation, which is called the three block case. We prove that the three block
theories all obey the BMW algebra.  We exemplify this result by treating in detail
the $SU(2)$ $2\times 2$ fused models, and showing explicitly the BMW structure.
We use the connection between the construction of solvable lattice models and conformal field theory.
This result is important to the solution of IRF lattice models and the development of new
models, as well as to knot theory.
\endpage 

\section{Introduction}

Solvable lattice models in two dimensions play a major role as models of phase transitions.
For a review see Baxter's book \REF\Baxter{R.J. Baxter, ``Exactly solved models in statistical
mechanics'', Academic Press, London, England, 1982.}\r\Baxter. They also have a
beautiful role in mathematics. For example, they allow for the definition of knot invariants,
which are crucial in the classification of knots. For a review see \REF\Wadati{M. Wadati, T. Deguchi,
and Y. Akutsu, Physics Reports 180 (4) (1989) 247.}\r\Wadati, and ref. therein.

This paper is concerned with Interaction Round the Face (IRF) solvable lattice models.
The first examples are the Ising and the hard hexagon models. 
These have been vastly generalized, see, e.g. \r\Wadati. In particular, in ref. \REF\Found{D. Gepner, arXiv: 
hep-th/9211100v2 (1992).}\r\Found, the construction of such  models from conformal field theory
was described. The conformal field theories in two dimensions  were first suggested by BPZ \REF\BPZ{A.A. Belavin, A.M. Polyakov and
A.B. Zamolodchikov, Nucl.Phys. B241 (1984) 393.}\r\BPZ,  constructing  the so called minimal models.
They were subsequently extended to WZW affine conformal field theories  \REF\Witten{E. Witten, Comm. Math.
Phys. 92 (1984) 455.}
\REF\KZ{V.G. Knizhnik and A.B. Zamolodchikov, Nucl. Phys. B247 (1984) 83.}
\REF\GW{D. Gepner and E. Witten, Nucl.Phys. B278 (1986) 493.}
\r{\Witten,\KZ,\GW}. 
The main examples of solvable IRF models, as discussed here, are related to  the WZW theories.

The algebraic structure of IRF models plays a significant role in their solution. It is also 
important in applications to knot theory. The face transfer matrix obeys some polynomial equation,
the order of which we call the number of blocks. 
This is further elucidated in the following discussion.
For two blocks it is known that the algebra is
Temperly--Lieb algebra \REF\TL{N. Temperly and E. Lieb, Proc.R.Soc. A 322 (1971) 251.}
\r\TL, yielding the, so called, graph--state IRF models.

Our aim here is to treat the three block case. We prove that three block IRF models obey the
Birman--Murakami--Wenzl (BMW) algebra, which was suggested in connection to knot theory
\REF\BW{J.S. Birman and H. Wenzl, Trans.Am.Math.Soc. 313 (1) (1989) 313.}
\REF\Murakami{J. Murakami, Osaka J.Math. 24 (4) (1987) 745.}\r{\BW,\Murakami}.
This is our main result. We exemplify by showing that the fused $2\times 2$ $SU(2)$ models
obey the BMW algebra.

These results are important to the understanding of IRF models, the construction of new ones,
and is mathematically interesting, particularly  in knot theory. We hope to further the understanding
the algebraic structure of IRF models, in general.

\section{Interaction round the face lattice models}
Fusion IRF models were constructed in ref. \r\Found.
We consider a square lattice.
Fix a rational conformal field theory $\cal O$. We denote the primary fields by $[p]$ where 
the unit field is denoted by $[1]$. On the sites of the lattice models we have some primary fields.
The interaction is around the faces of the model.
The partition function of the model is given by
\def\om{\omega \left (\matrix {  a & b \cr c & d \cr}\bigg | u\right )}
$$Z= \sum_{\rm configurations} \prod_{\rm faces} \om,\e$$
where the sum is over all primary fields sitting on the sites and 
$$\om \e$$
is some Boltzmann weight to be defined. Here $u$ is some parameter called the spectral parameter.
 
 Denote by $f_{a,b}^c$ the fusion coefficient of the conformal field theory $\cal O$, where 
 $[a]$, $[b]$ and $[c]$ are some primary fields. We fix a pair of primary fields, $[h]$ and $[v]$. The model is defined
 by the admissibility condition that the Boltzmann weights vanishes unless the fields around 
 the face  obey  \r\Found,
 $$f_{h b}^a f_{vd}^b f_{hd}^c f_{v c}^a>0.\e$$ 
 Accordingly, we denote this model as ${\rm IRF}({\cal O},h,v)$.  
 We find it convenient to define the face transfer matrix, through its matrix elements,
 $$\left<  a_1,a_2,\ldots,a_n | X_i(u) | a^\pr_1, a^\pr_2,\ldots, a^\pr_n \right>=\left[
 \prod_{j\neq i} \delta_{a_j,a_j^\pr} \right] \,\omega\left ( \matrix{ a_{i-1} & a_i\cr a_i^\pr & a_{i+1}\cr}\bigg | u\right).\e$$
 In this language, the conditions for the solvability of the model, which is the celebrated Yang Baxter 
 equation becomes,
 $$X_i(u) X_{i+1} (u+v) X_i(v)=  X_{i+1}(v) X_i(u+v) X_{i+1}(u),\e$$
 $$X_i(u) X_j(v)=X_j(v) X_i(u),\quad{\rm for}\ |i-j|\geq2.\e$$
 The Yang Baxter equation implies that the transfer matrices for all values of  $u$ commute with one another.
 
 At the limit $u\rarrow i \infty $ the Yang Baxter equation (YBE) becomes,
 $$X_i X_{i+1} X_i=X_{i+1} X_i X_{i+1},\quad {\rm where }\quad X_i=\lim_{u\rarrow i \infty} e^{i(n-1) u} X_i(u).\e$$
 $$X_i X_j=X_j X_i \quad {\rm where }\quad |i-j|\geq 2.\e$$
 This is the braid group relation. $n$ is the number of blocks. The exponent
 pre--factor is necessary to make the limit finite as is discussed below.
 
 We have a natural candidate for the braid group which obeys the admissibility condition eq. (3). This 
 is the braiding matrix of the conformal field theory $\cal O$, which expresses the braiding of the conformal blocks
 \REF\Braid{G. Moore and N. Seiberg, Phys. Lett. B 212 (1988) 451.}\r\Braid. This matrix is denoted by $C$ and it obeys the Hexagon
 relation \r\Braid, which for $h=v$ is equivalent to the braid group relations, eq. (7--8). We define
 $$\lim_{u\rarrow i\infty} e^{i(n-1)u} \om=C_{c,d}\left[ \matrix{h &v\cr a& b \cr } \right ].\e$$ 
 With this definition the Boltzmann weights $\omega$ obey the admissibility condition for the lattice 
 model ${\rm IRF} ({\cal O},h,v)$, and the face matrix, $X_i$, obeys the braid group relations, eqs. (7--8).

 Denote by $\Delta_p$ the dimension of the primary field $[p]$. 
 Assume also that $h=v$ and 
 $$[h]\times [h]=\sum_{i=0}^{n-1} \psi_i,\e$$
 where the product is in terms of the fusion rules,  and $\psi_i$ are some primary fields. 
 We denote by $\Delta_i$ the dimension of the primary field $\psi_i$, and
 $n$ is
 the number of primary fields in the product. 
 We call this the $n$ block case. 
 The face operator  $X_i$, which is equal to the
 braiding matrix, has then the eigenvalues,
 $$\lambda_i=\epsilon_i e^{i\pi(\Delta_h+\Delta_v-\Delta_i)},\e$$
 where $\epsilon_i=\pm1$, according to whether the product is symmetric or anti--symmetric. We
 shall assume that $\epsilon_i=(-1)^i$.
 
 The matrix $X_i$ then is seen to obey an $n$th order
 polynomial equation,
 $$\prod_{p=0}^{n-1} (X_i-\lambda_p)=0.\e$$
 
 The fact that $X_i$ obeys an $n$th order equation allows us to define projection operators.
 We define,
 $$P_i^a=\prod_{p\neq a} \left[ {X_i-\lambda_p\over \lambda_a-\lambda_p}\right ].\e$$
 The projection operators obey the relations,
 $$\manyeq{\sum_{a=0} ^{n-1} P_i^a&=1\cr
                    P_i^a P_i^b&=\delta_{ab} P_i^a\cr
                    \sum_{a=0}^{n-1} \lambda_a P_i^a&=X_i \cr} $$
                    
 We shall assume that the theory is real. This implies that $h=\bar h$. So in the fusion rules
 $[h]\times [h] =1+\ldots$, or $\psi_0=[1]$. In ref. \r\Found, a conjecture for a trigonometric solutions of the
 YBE, eqs. (5--6), was proposed. For this purpose we introduce the parameters,
 $$\zeta_i=\pi (\Delta_{i+1}-\Delta_i)/2,\e$$
 where $\Delta_i$ is the dimension of the field $\psi_i$, and $\Delta_0=0$. The trigonometric solution for the
 Yang Baxter equation is then
 $$X_i(u)=\sum_{a=0}^{n-1} f_a(u) P_i^a,\e$$
 where the functions $f_a(u)$ are defined by
 $$f_a(u)=\left[ \prod_{r=1}^{a} \sin(\zeta_{r-1}-u)\right ] \left [ \prod_{r=a+1}^{n-1} \sin(\zeta_{r-1}+u)\right]\bigg/\left[\prod_{r=1}^{n-1} \sin(\zeta_{r-1})\right].\e$$
 
In many cases  it was  verified explicitly that the  conjectured Boltzmann weights, eqs. (16--17),  obey
the YBE equation. For a review see, e.g.,  ref.  \r\Wadati. In what follows we shall assume that 
for our  conjectured Boltzmann weights, the YBE  indeed holds. It appears that all solvable IRF
 models can be derived in this way from some conformal field theory, provided that they
 have a second order phase transition.
 
The Boltzmann weights, $X_i(u)$ obey several properties \r\Found. The first one is crossing,
$$\omega\left(\matrix{a & b\cr c&d\cr} \bigg| \lambda-u\right)=\left[ {S_{b,0} S_{c,0}\over S_{a,0} S_{d,0}}\right] ^{1/2}\omega\left(\matrix{c& a\cr d &b\cr}\bigg | u\right),\e$$
where $S_{a,b}$ is the modular matrix, and $\lambda=\zeta_0$ is the crossing parameter.

Define the element of the algebra,
$$E_i=X_i(\lambda).\e$$
Since $X_i(0)=1$ we find from the crossing relation an expression for $E_i$,
$$E \left(\matrix{ a& b\cr c& d\cr} \right)=\left[{S_{b,0} S_{c,0} \over S_{a,0} S_{d,0}}\right] ^{1/2}\delta_{a,d}.\e$$
From this expression it easy to verify that $E_i$ obeys the Temperly--Lieb algebra,
$$E_i E_{i\pm 1} E_i=E_i.\e$$
For the square of $E_i$ we find,
$$E_i^2=\beta E_i,\quad {\rm where}\quad  \beta=\prod_{r=1}^{n-1} {\sin(\lambda+\zeta_{r-1})\over \sin(\zeta_{r-1})}.\e$$
In the case of two blocks $n=2$ this solves completely for $X_i(u)$ showing that the solution is
a graph state IRF \r\Found. We conclude that for two blocks the algebra is Temperly--Lieb. It is
noteworthy that for any number of blocks, $E_i$ obeys the Temperly--Lieb algebra.

Another relation which is evident is the unitarity,
$$X_i(u) X_i(-u)=\rho(u) \rho(-u) 1_i ,\e$$
where 
$$\rho(u)=\prod_{r=1}^{n-1} {\sin(\zeta_{r-1}-u)\over \sin(\zeta_{r-1})}.\e$$

\section{Birman Murakami and Wenzl algebra}

Our purpose is to generalize the two blocks result to three blocks. Our main result is that
every three block theory obeys the Birman--Murakami--Wenzl algebra (BMW) \r{\BW,\Murakami}. First, for this purpose, we list the relationships of the algebra.
There are two generators of the algebra, $G_i$ and $E_i$.
The relations are,
$$G_i G_j=G_j G_i {\ \rm  if\ }  |i-j|\geq 2,$$
$$G_i G_{i+1} G_i=G_{i+1} G_i G_{i+1},\qquad E_i E_{i\pm1} E_i=E_i,$$
$$G_i-G_i^{-1} =m(1-E_i),$$
$$G_{i\pm1} G_i E_{i\pm1}=E_i G_{i\pm1} G_i=E_i E_{i\pm1},\qquad G_{i\pm1} E_i G_{i\pm1}=
G_i^{-1} E_{i\pm1} G_i^{-1},$$
$$G_{i\pm1 } E_i E_{i\pm1}=G_i^{-1} E_{i\pm1},\qquad E_{i\pm1} E_i G_{i\pm1}=E_{i\pm1} G_i^{-1},$$
$$G_i E_i=E_i G_i=l^{-1} E_i,\qquad E_i G_{i\pm1} E_i=l E_i.\e$$
These relations imply the additional relations,
$$E_i E_j=E_j E_i {\ \rm if \ } |i-j|\geq 2,\qquad (E_i)^2=[(l-l^{-1})/m+1] E_i.\e$$
Here, $l$ and $m$ are the two parameters of the algebra. The BMW algebra is  defined  here according to Kauffmann's 
`Dubrovnik' version of the algebra \REF\Kauffmann{L. Kauffmann, ``Knot Theory", World Scientific,
Singapore (1991).}\r\Kauffmann, which is used in the defenition of Kauffmann's polynomial.

The BMW algebra is known to have a canonical Baxterization \r{\BW,\Murakami}.
This is given by two parameters $\lambda$ and $\mu$. The parameters are related to $m$ and $l$,
as will be described below. The face operator is defined by,
$$U_i(u)=1-{i \sin (u)\over 2 \sin(\lambda) \sin(\mu)} \left [ e^{-i (u-\lambda)} G_i-e^{i(u-\lambda)} G_i^{-1} \right ].\e$$
Here $1$ stands for the unit matrix.
With the definition of the BMW algebra, eqs. (25--26), $U_i(u)$ obeys the Yang Baxter equation,
$$ U_i(u) U_{i+1} (u+v) U_i (v)=U_{i+1} (v) U_i(u+v) U_{i+1} (u).\e$$ 

We shall concentrate now on the three block case, $n=3$. In this case, the solution for the
face transfer matrix, eqs. (16--17), becomes,
$$\twoline{X_i(u)=\big[ P^0_i \sin(\zeta_0+u) \sin(\zeta_1+u)+P^1_i \sin(\zeta_0-u) \sin(\zeta_1+u)+}{P^2_i 
\sin(\zeta_0-u)\sin(\zeta_1-u)  \big] /\left[ \sin(\zeta_0)\sin(\zeta_1)\right].}$$
We also define
$$X^t_i=\lim_{u\rarrow -i\infty} e^{-2 i u}  X_i(u).\e$$

According to the unitarity condition, eqs. (23--24), 
$X_i X^t_i$ is proportional to one. We wish to scale them, so that they are the inverse of one another.
The  scale constant can be read from eqs. (23--24), the unitarity condition,
$$X_i X_i^t=\bigg(\lim_{u\rarrow i\infty} e^{2 i u} \rho(u)  \bigg) \bigg( \lim_{u\rarrow -i\infty } e^{-2 i u}
\rho(u)\bigg)=1/w^2,\e$$
where
$$w=4 \sin(\zeta_0)\sin(\zeta_1),\e$$
is the necessary scale constant. 

Our purpose now is to connect our solution to the YBE $X_i(u)$, eqs. (19), with algebraic
solution $U_i(u)$, eq. (27),
$$U_i(u)=P_i^0+P_i^1+P_i^2-{i \sin(u)\over 2\sin(\lambda)\sin(\mu)} \left(e^{-i(u-\lambda)} G_i-
e^{i(u-\lambda)} G_i^{-1} \right).\e$$  
To do this, we identify $\zeta_0$ with $\lambda$ (as before) and $\zeta_1$ with $\mu$.
We also identify the generators $G_i$ and $G_i^{-1}$ to be proportional to $X_i$ and
$X_i^t$ respectively,
$$\manyeq{G_i&=4\sin(\lambda)\sin(\mu) e^{-i\lambda} X_i,\cr
                    G_i^{-1}&=4 \sin(\lambda)\sin(\mu) e^{i\lambda} X_i^t.\cr} $$
The phase is arbitrary, and we fixed it
to be compatible with the BMW algebra, eqs. (25--26). Indeed, from eq. (31), we have
$G_i G_i^{-1}=1$, as it should. We identify $G_i$ as the generator of the BMW algebra, eqs. (25--26),
along with $E_i=X_i(\lambda)$. Our purpose is to show that with this definition the BMW algebra
is indeed obeyed.

We substitute the scale definition, eq. (34), into the expression for the Baxterized BMW algebra, $U_i(u)$, eq. (33).
We find after some algebra that $X_i(u)$ and $U_i(u)$ are identical,
$$U_i(u)=X_i(u),\e$$
for any $P_i^0$, $P_i^1$ and $P_i^2$, provided only that they obey $P_i^0+P_i^1+P_i^2=1$, which is 
true.

We conjectured that $X_i(u)$ obeys the Yang Baxter equation, eqs. (5--6). This
was proved for  many cases. 
It then follows that $U_i(u)$ obeys  the YBE also. By expanding the face weight, as 
in eq. (33), this, in turn, implies that $G_i$ and $E_i$ obey the Birman--Murakami--Wenzl
algebra, eqs. (25--26), which is the main result of this paper. We basically, invert the logic:
instead of proving that $U_i(u)$ obeys the Yang Baxter equation, from the BMW relations,
we prove the BMW relations from the Yang Baxter equation for $U_i(u)$. 

We get now to determining the parameters $l$ and $m$. We have  the equation, eq. (25),
$$G_i-G_i^{-1}=m(1-E_i),\e$$
where $E_i=U_i(\lambda)$. This is the Skein relation. In particular $E_i$ obeys the Temperly--Lieb algebra,  eqs. (21--22),
which is part of the BMW algebra, eqs. (25--26). By calculating $U_i(\lambda)$, from eq. (27), we find
$$m=-2 i \sin(\mu).\e$$
By calculating the equation,
$$G_i E_i=E_i G_i=l^{-1} E_i,\e$$
we find that 
$$l=-e^{i(2\lambda+\mu)}.\e$$
Finally, we had before that $E_i^2=\beta E_i$, eq. (22), where
$$\beta={\sin(2\lambda)\sin(\lambda+\mu)\over \sin(\lambda)\sin(\mu)}.\e$$
We know from the BMW algebra that we should have,
$$\beta=(l-l^{-1})/m+1.\e$$
Substituting $l$ and $m$ we find that this relation is indeed obeyed.

\section{The fused $SU(2)$ model.}

Let us move now to a concrete example. This is the model ${\rm IRF}(SU(2)_k,[2],[2])$. Namely,
the conformal field theory $SU(2)_k$, which is a WZW affine theory, at level $k$, with the admissibility fields $h=v=[2]$. We denote by $l$ the weights of
$SU(2)_k$ representations, which are  twice the isospin, $l=0,1,2,\ldots,k$. So $[2]$ is the field with isospin one.

The fields appearing in the product $hv$ are
$$[2]\times  [2]=[0]+[2]+[4].\e$$
The dimensions of the fields in the theory are given by
$$\Delta_l={l(l+2)\over 4 (k+2)}.\e$$
So we find for $\zeta_0$ and $\zeta_1$ the values, according to eq. (15),
$$\zeta_0=\lambda={\pi\over 2}{ 2\cdot 4\over 4(k+2)}={\pi\over k+2},\e$$
and 
$$\zeta_1={\pi\over 2} {4\cdot 6-2\cdot 4\over 4(k+2)}={2\pi\over k+2}=2 \lambda.\e$$

Thus the face transfer matrix, $X_i(u)$, is given by eq. (29), with the substitution of the above 
values of $\zeta_0$ and $\zeta_1$. In fact, the exact same weights were given  in a
paper by Pasquier \REF\Pasquier{V. Pasquier, Comm.Math.Phys. 118 (1988) 355.}\r\Pasquier.
In this paper it is proved that $X_i(u)$ obeys the Yang Baxter equation, with the appropriate choice
of $P^0_i$ and $P^1_i$. It thus follows, according from the discussion above, that $G_i$ and
$E_i$ obeys the BMW algebra, with $\lambda=\pi/(k+2)$ and $\mu=2\lambda$.
The values of $l$ and $m$ for this model are,
$$l=-e^{4 i\lambda},\qquad m=-2 i\sin(2\lambda).\e$$

Explicit expressions for all the Boltzmann weights of the model are listed in the appendix.
We checked numerically that all the relations of the BMW algebra, eqs. (25--26), are obeyed for
this model.
We defined $G_i$ and $G_i^{-1}$  as in eq. (34).
We checked for levels $k=8,9,10,12$ and  we find a complete agreement with
the BMW algebra, with all the relations fulfilled. 
We conclude that this model obeys the BMW algebra.

\section{Discussion}
In this paper we proved that any three block IRF model obeys the BMW algebra, along with
its Baxterization. We treated explicitly the fused model IRF$(SU(2)_k,[2],[2])$,
and showed that it obeys the BMW algebra, also by direct computations.

Other important models are the $BCD$ IRF models constructed by Jimbo et. al. 
\REF\BCD{M. Jimbo, T. Miwa and M. Okado, Comm.Math.Phys. 116 (1988) 353.}\r\BCD.
These models can be described as IRF$(G,[v],[v])$, where $G$ is either the $B_n$,
$C_n$ or $D_n$ WZW theory at level $k$, and $[v]$ stands for	 the vector representation.
Since $[v]\times [v]$ contains three fields, this is a three block model. Thus, our analysis  in this
paper is applicable to these models, proving that they obey the BMW algebra.

Other three block theories were described in ref. \REF\Dov{R. Dovgard and D. Gepner, 
J.Phys. A42 (2009) 304009.}\r\Dov, at the level of the conformal data, as theories
with low number of primary fields. It will be interesting to study the IRF models associated 
with these models, as well. In particular, from our analysis we expect them to obey the BMW
algebra.

An important question, left to the future, is to find the algebras corresponding to IRF models with more than
three blocks. We believe that our methods can be useful to study these models, as well.
Some of the relations of the general models are known \r\Found, already. Finding these algebras will be 
of considerable importance to knot theory.

\ack
It is our pleasure to thank Ida Deichaite for much discussions and encouragement.

\def\Wtwo#1#2#3#4#5{{\omega\left( \matrix{#4 & #1\cr #3 & #2 \cr}\bigg | #5  \right)}}
\def\frac #1 #2 {#1\over #2}
\appendix
\hbox{\bf Weights of $2\times 2$ fused model.\hfill}
These are the weights of the model IRF(SU(2),[2],[2]). We define 
$$\lambda={\pi\over k+2}\e $$
and
$$s[x]={\sin(x)\over \sin(\lambda)}.\e$$
The weights are taken from ref. \REF\Pearce{E. Tartaglia and P. Pearce, J. Physics A49 (2016)  18.}\r\Pearce. The weights were originally computed in ref. \REF\Jimbo{E. Date, M. Jimbo, T. Miwa and
M. Okado, Lett. in Math. Phys. 12 (1986) 209.}\r\Jimbo, by the fusion procedure.
We shifted the primary field $a\rarrow a+1$ so the weights range over
$a=1,2,\ldots, k+1$,  giving  the dimension of the representation.

The $19$ Boltzmann weights of the model are as follows:

$$\manyeq{
\Wtwo{a}{a\mp2}{a}{a\pm2}{u} &= {s(u-2 \lambda ) s(u-\lambda ) \over s(2 \lambda )} \cr
\Wtwo{a}{a\pm2}{a}{a}{u} &=
\Wtwo{a}{a}{a}{a\pm2}{u} = -{s(u-\lambda ) s((a\pm1) \lambda \mp u) \over s((a\pm1) \lambda )} \cr
\Wtwo{a\pm2}{a}{a}{a}{u} &= -{s((a\mp1) \lambda )s(u)  s(a \lambda \pm u) \over s(2 \lambda ) s(a \lambda ) s((a\pm1)\lambda )} \cr
\Wtwo{a}{a}{a\pm2}{a}{u} &= -{s(2 \lambda ) s((a\pm2) \lambda )s(u) s(a \lambda \pm u)\over s((a-1) \lambda ) s((a+1) \lambda )}\cr
\Wtwo{a\mp2}{a}{a\pm2}{a}{u} &= {s((a\mp2) \lambda ) s((a\mp1) \lambda ) s(u)s(\lambda +u) \over s(2 \lambda )   s(a \lambda ) s((a\pm1) \lambda )}\cr}$$
$$\manyeq{
\Wtwo{a\pm2}{a}{a\pm2}{a}{u} &= {s(a \lambda \pm u) s((a\pm1) \lambda \pm u) \over s(a \lambda ) s((a \pm 1) \lambda )}\cr
\Wtwo{a}{a}{a}{a}{u} &={s(a\lambda\pm u)s((a\pm 1)\lambda\mp u) \over s(a\lambda)s((a\pm1)\lambda)}+{s((a\pm 1)\lambda)s((a\mp2)\lambda)s(u)s(u-\lambda) \over s(2\lambda)s(a\lambda)s((a\mp1)\lambda)}\cr
\Wtwo{a\pm2}{a\pm2}{a}{a}{u} &=\Wtwo{a\pm2}{a}{a}{a\pm2}{u}= {s((a\pm3) \lambda ) s(u) s(u-\lambda ) \over s(2 \lambda ) s((a\pm1) \lambda )} \cr
}$$

\refout
 \bye

%% file: mydef.tex
\def\e{\adveq\eqno{\rm (\chapterlabel\the\equanumber)}}

\def\adveq{\global\advance\equanumber by 1}
\def\myeq{{\rm \chapterlabel\the\equanumber}}
\def\rarrow{\rightarrow}

\def\twoline#1#2{\displaylines{\qquad#1\hfill(\adveq\myeq)\cr\hfill#2
\qquad\cr}}

\def\manyeq#1{\eqalign{#1}\e}

\def\semidirect{\mathrel{\raise0.04cm\hbox{${\scriptscriptstyle |\!}$
\hskip-0.175cm}\times}}


\def\ref#1{$^{[#1]}$}

\def\pr#1{#1^\prime}
 
\def\r#1{$[\rm#1]$}

\def\e{\adveq\eqno{\rm (\chapterlabel\the\equanumber)}}

\def\adveq{\global\advance\equanumber by 1}
\def\myeq{{\rm \chapterlabel\the\equanumber}}
\def\rarrow{\rightarrow}

\def\twoline#1#2{\displaylines{\qquad#1\hfill(\adveq\myeq)\cr\hfill#2
\qquad\cr}}

\def\manyeq#1{\eqalign{#1}\e}

\def\semidirect{\mathrel{\raise0.04cm\hbox{${\scriptscriptstyle |\!}$
\hskip-0.175cm}\times}}


\def\ref#1{$^{[#1]}$}

\def\pr#1{#1^\prime}
 
\def\r#1{$[\rm#1]$}